\documentclass[superscriptaddress,reprint,prb,showpacs]{revtex4-1}
\usepackage{float}
\usepackage[utf8]{inputenc}
\usepackage{times}
\usepackage{graphicx}
\usepackage{url}
\usepackage{amsmath}
\usepackage{amsfonts}
\usepackage{amssymb}
\usepackage[mathscr]{euscript}
\usepackage{cancel}
\usepackage{color}
\usepackage{dcolumn}

\begin{document}

	\author{Patrick P. Hofer}
	\affiliation{Department of Physics, McGill University, 3600 rue University, Montreal, Quebec, H3A 2T8, Canada }
	\affiliation{D{\'e}partement de Physique Th{\'e}orique, Universit{\'e} de Gen{\`e}ve, 1211 Gen{\`e}ve, Switzerland }
	\author{J.-R. Souquet}
	\affiliation{Department of Physics, McGill University, 3600 rue University, Montreal, Quebec, H3A 2T8, Canada }
	\author{A.\,A. Clerk}
	\affiliation{Department of Physics, McGill University, 3600 rue University, Montreal, Quebec, H3A 2T8, Canada }
	\date{\today}

\title{Quantum heat engine based on photon-assisted Cooper pair tunneling}
	\begin{abstract}
We propose and analyze a simple mesoscopic quantum heat engine that exhibits both high-power and high-efficiency.  The system consists of a biased Josephson junction coupled
to two microwave cavities, with each cavity coupled to a thermal bath.  Resonant Cooper pair tunneling occurs with the exchange of photons between cavities, and a temperature difference
between the baths can naturally lead to a current against the voltage, and hence work.  As a consequence of the unique properties of Cooper-pair tunneling,
the heat current is completely separated from the charge current.  This combined with the strong energy-selectivity of the process leads to an extremely high efficiency.
	\end{abstract}
\maketitle

\textit{Introduction.} Quantum heat engines are devices that convert heat into work which are described by the laws of quantum mechanics. In addition to possible energy-harvesting and
refrigeration applications, these devices provide a testbed for the study of thermodynamics in quantum systems.\cite{giazotto:2006,benenti:2013} Several proposals for thermoelectric heat engines have been put forward including systems based on quantum dots\cite{jordan:2013,sothmann:2015} and coherent interferometers.\cite{hofer:2015}
Approaches using a heat source which does not exchange particles with the conductor \cite{rutten:2009,entin:2010,sanchez:2011,entin:2012,ruokola:2012,jiang:2012,sothmann:2012,sothmann:2012epl,jordan:2013,sothmann:2013,jiang:2013,jiang:2013prb,donsa:2014,mazza:2014,choi:2015,entin:2015,whitney:2015,thierschmann:2015,roche:2015,hartmann:2015} are attractive, as they more easily allow the establishment of a thermal gradient.
Of these, proposals using microwave photons to mediate the heat flux \cite{ruokola:2012,bergenfeldt:2014,henriet:2015} are particularly promising, as they allow for a macroscopic spatial separation of hot and cold reservoirs, and because of the rapid progress in relevant experiments on hybrid cavity-conductor systems.\cite{masluk:2012,altimiras:2014}

\begin{figure}[t!]
\centering
\includegraphics[width=.9\columnwidth]{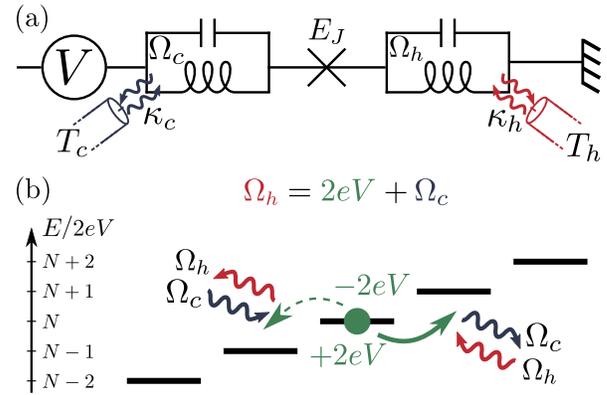}
\caption{Setup and sketch of the heat conversion process. (a) The heat engine consists of a voltage biased Josephson junction coupled to two cavities (frequencies $\Omega_h>\Omega_c$), each coupled via a transmission line to their own thermal environment (temperatures $T_h$ and $T_c$,  damping rates $\kappa_h$ and $\kappa_c$). (b) The bias voltage is chosen as $2eV=\Omega_h-\Omega_c$. If the occupation probability of the hot cavity (subscript $h$) is higher than the occupation probability of the cold cavity (subscript $c$), Cooper pairs tunnel against the bias annihilating a photon in the hot and creating a photon in the cold cavity. Thus a heat current from the hot to the cold cavity is converted into a supercurrent against the voltage bias, providing work. Here $N$ denotes the number of Cooper pairs that have tunneled against the bias.}
  \label{fig:schematics}
\end{figure}

In this work, we present a new kind of high-power and high-efficiency mesoscopic heat engine which also makes use of microwave photons.  Unlike previous proposals, we use a superconducting
electronic system, consisting of a voltage-biased Josephson junction coupled to two resonant cavities,  see Fig.~\ref{fig:schematics}.
In spite of an extensive literature on the thermal properties of hybrid superconducting-normal structures,\cite{claughton:1996,eom:1998,hwang:2015,virtanen:2007,chandrasekhar:2009}
superconductors have received relatively little attention as building blocks for heat engines due to their particle-hole symmetry. Only recently has the possibility of using a superconducting contact to separate charge from heat current in a thermoelectric engine been put forward.\cite{mazza:2015}

In our proposed engine, the work is provided by Cooper pairs tunneling against a voltage bias while the heat current is mediated by two cavities coupling to baths at different temperatures.
The key ingredient in our proposal is the fact that the Cooper pairs carry no entropy (and thus no heat) which allows for a complete separation of the charge and heat currents. As a consequence, the conversion of heat into work happens with a high and universal efficiency beating the Curzon-Ahlborn bound\cite{curzon:1975} at maximum power. Although it is well known that this bound does not constitute an exact bound,\cite{broeck:2005,esposito:2010,benenti:2013} it nevertheless serves as a useful reference value. Furthermore, using microwave cavities, the hot and the cold reservoir can be separated by macroscopic distances in our proposal (e.g. they can be located at the end of transmission lines), facilitating the task of maintaining a heat gradient across the sample.

\begin{table*}
\def\arraystretch{1.3}
\begin{tabular}{|c|c|c|c|c|c|c|c|c|c| }
\hline
  \,$\Omega_h/2\pi$\, & \,$\Omega_c/2\pi$\, & \,$\kappa/2\pi$\, &\, $E_J/2\pi$\, &
  \, $T_h$ \,&\, $T_c$\,&\, $\lambda$\, &\, $I$\,&\,$P$ \, &\, $\eta$\,\\\hline\hline
     \,$13.5$\,GHZ\, &\, $3$\,GHZ\, &\, $0.06$\,GHz\, & \,$0.3\,$GHz =$1.24/2\pi\,\mu$eV \,&
     \, $0.93$\,K\, & \,$0.06\,$K\,& \,$0.36$\, &\,$23\,$pA \,&\,$0.5\,$fW \,&\, $77.8\,$\%\,\\
\hline
\end{tabular}
\caption{Realistic Parameters for operating the proposed heat engine. Here $\kappa=\kappa_h=\kappa_c$ and $\lambda=\lambda_h=\lambda_c$.}
\label{tab:params}
\end{table*}

Before discussing our system in detail, it is worthwhile to discuss how it differs from previous works.
Although our proposal relies on cavities with reasonably high zero-point phase fluctuations, we note that in contrast to Ref.~\onlinecite{henriet:2015}, no asymmetry in the zero-point fluctuations is required. The direction of current in our system is fully determined by the bias voltage and the photon occupation numbers in the cavities. Similarly to Ref.~\onlinecite{bergenfeldt:2014}, we make use of microwave cavities to spatially separate the thermal baths. However, in our proposal the heat baths are not provided by the electronic contacts, thus allowing for the complete separation of the heat flow from the electric current. In analogy to the energy selectivity of a Josephson junction, previous works investigated the use of quantum dots as energy filters to obtain an efficient conversion from heat to work.\cite{sanchez:2011,ruokola:2012} In contrast to our proposal, these works rely on highly asymmetric tunneling rates and the efficiency at maximum power remains below the Curzon-Ahlborn bound. Finally, in Ref.~\onlinecite{mazza:2015} a superconducting terminal is used to separate the heat from the charge current. However, the thermal baths in this setup are provided by electronic leads and thus lack the highly peaked spectral density of our proposal.

\textit{Setup and model.} The two cavities involved in our proposal have different frequencies $\Omega_h>\Omega_c$ and are coupled to baths at different temperatures such that their mean occupation numbers fulfill $\langle\hat{n}_h\rangle>\langle\hat{n}_c\rangle$. Here and below the index $h$ ($c$) denotes the hot (cold) cavity. The heat baths are characterized by an occupation number $n_B^\alpha=[\exp({\frac{\Omega_\alpha}{k_BT_\alpha}})-1]^{-1}$ which depends on the cavity frequency $\Omega_\alpha$ and the bath temperature $T_\alpha$.
Due to a voltage bias across the Josephson junction, a Cooper pair can cross the junction only by exchanging energy with the photonic degrees of freedom. For a bias voltage of the form ($\hbar=1$ throughout this paper)
\begin{equation}
\label{eq:voltage}
2eV=k\Omega_h-l\Omega_c,
\end{equation}
where $k$ and $l$ are integers, a Cooper pair can tunnel against the bias by absorbing $k$ photons from the hot cavity and emitting $l$ photons to the cold cavity.
This is the process by which the proposed heat engine converts heat into work. The hot bath provides the heat $k\Omega_h$, carried by the absorbed photons, which is converted into the work $2eV$, represented by the energy gain of the Cooper pair. The excess heat $l\Omega_c$ is dumped into the cold bath and is carried by the emitted photons. The efficiency of converting heat into work for a single Cooper pair tunneling against the bias thus reads
\begin{equation}
\label{eq:eta}
\eta=\frac{2eV}{k\Omega_h}=1-\frac{l\Omega_c}{k\Omega_h}.
\end{equation}
As shown below, this coincides with the efficiency for a realistic heat engine obtained from a rigorous calculation.
In the above discussion, we neglected Cooper pairs tunneling with the bias which provide negative work. These processes also contribute a negative heat current (i.e.~they are accompanied by a heat flux from the cold to the hot cavity).  Including them reduces the net work and heat input by the same factor, leaving the efficiency unchanged. The reason for the high and universal efficiency
we obtain is ultimately rooted in the sharp energy-filtering feature of the Josephson junction.

\begin{figure*}[t!]
\centering
\includegraphics[width=.9\textwidth]{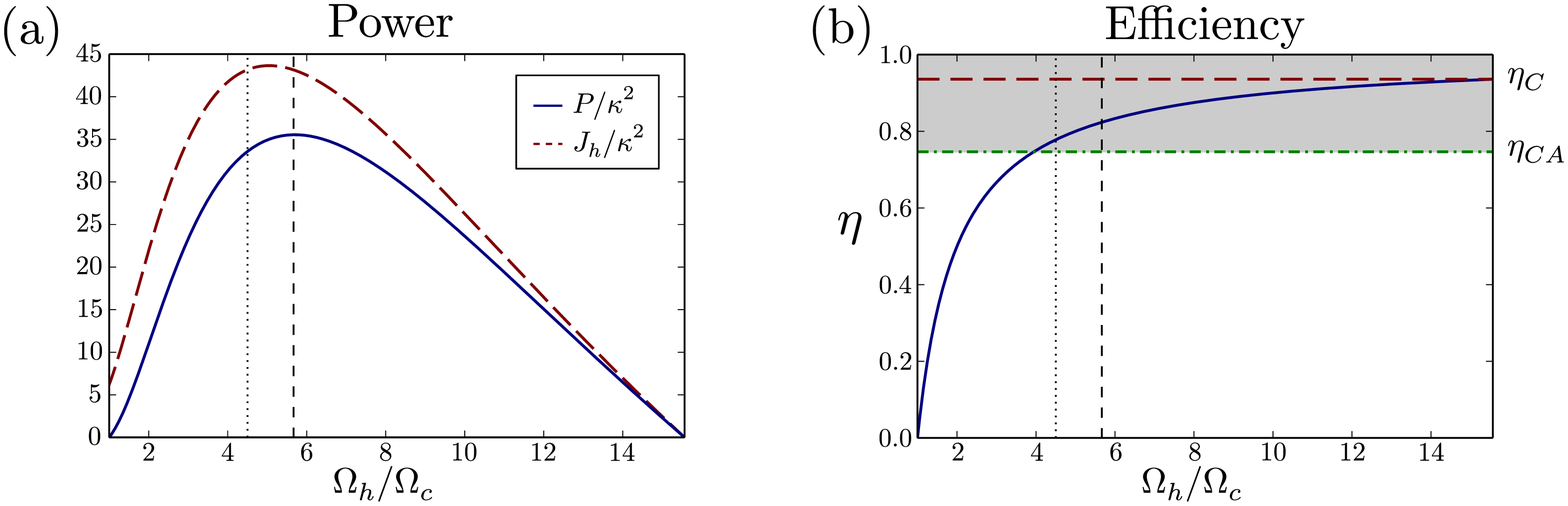}
\caption{Power and efficiency as a function of the voltage. (a) Power (solid, blue) and heat current (dashed, red), (b) efficiency as a function of $\Omega_h$ while keeping $T_h$ and $\Omega_c$ constant. At $\Omega_h=\Omega_c$, the power and the efficiency vanish because the voltage vanishes. At the stopping voltage $\Omega_h/T_h=\Omega_c/T_c$ (right end of plots), the charge and heat currents vanish and Carnot efficiency $\eta_C=1-\frac{T_c}{T_h}$ is reached. The dashed vertical lines correspond to maximal power. Note that at this point the efficiency is above the Curzon-Ahlborn bound $\eta_{CA}=1-\sqrt{\frac{T_c}{T_h}}$. The dotted vertical lines at $\Omega_h=4.5\Omega_c$, correspond to the performance analysis in the main text. Parameters: $E_J=5\kappa$, $\Omega_c=50\kappa$, $k_BT_h=325\kappa$, $n_B^c=0.1$ (corresponds to $k_BT_c=21\kappa$), $\lambda_h=\lambda_c=0.36$.}
  \label{fig:power}
\end{figure*}

While the efficiency in Eq.~\eqref{eq:eta} is independent of temperature, the magnitude and sign of the current (or, equivalently, the power) depends on temperature. As discussed below in more detail, we only get a net positive work (i.e.~net current against the bias) if $l\Omega_c/(k\Omega_h)>T_c/T_h$. This necessarily implies that Eq.~\eqref{eq:eta} is bounded by the Carnot efficiency $\eta_C=1-T_c/T_h$ which can be reached in the limit of vanishing output power [cf.\,Fig.\,\ref{fig:power}\,(b)]. In addition to these high and universal efficiencies, our heat engine achieves powers which compare well with existing proposals using hybrid microwave-cavities.\cite{bergenfeldt:2014} The generation of a dc current here is somewhat reminiscent of the well-known physics of Shapiro steps.\cite{shapiro:1963,shapiro:1964} Unlike Shapiro steps, there is no mean ac voltage driving the junction in our system: instead, the cavities only drive the junction with a noisy, narrow-band voltage.

For a more quantitative description, we start from the Hamiltonian\cite{armour:2013,gramich:2013}
\begin{equation}
\label{eq:hamiltonian}
\hat{H}=\Omega_h\hat{a}_h^\dagger\hat{a}_h+\Omega_c\hat{a}_c^\dagger\hat{a}_c-E_J\cos\left(2eVt+2\hat \varphi_h+2\hat \varphi_c\right),
\end{equation}
where the operator $\hat{a}_\alpha$ destroys a photon in the single-mode cavity $\alpha$ with frequency $\Omega_\alpha$. The phase of the Josephson junction is driven both by an external voltage and the flux $\hat \varphi_{\alpha}= \lambda_{\alpha}(\hat{a}_\alpha^\dagger+\hat{a}_\alpha)$ associated with each cavity. The coupling between the cavities and the Josephson junction depends on the amplitude of the cavity zero-point phase fluctuations $\lambda_{\alpha}=\sqrt{\pi e^2Z_\alpha/h }$  with $Z_\alpha$ being the impedance of cavity $\alpha$.
The current operator for this system reads
\begin{equation}
\label{eq:currop}
\hat{I}=-I_c\sin\left(2eVt+2\lambda_h(\hat{a}_h^\dagger+\hat{a}_h)+2\lambda_c(\hat{a}_c^\dagger+\hat{a}_c)\right),
\end{equation}
with the critical current $I_c=2eE_J$. Here a positive value implies a current against the bias.

Including the coupling to the heat baths, the dynamics is determined by the master equation
\begin{equation}
\label{eq:master}
\begin{aligned}
\partial_t\hat{\rho}=-i[\hat{H},\hat{\rho}]&+\kappa_h(n_B^h+1)\mathcal{D}[\hat{a}_h]\hat{\rho}+\kappa_hn_B^h\mathcal{D}[\hat{a}_h^\dag]\hat{\rho}\\&+\kappa_c(n_B^c+1)\mathcal{D}[\hat{a}_c]\hat{\rho}+\kappa_hn_B^c\mathcal{D}[\hat{a}_c^\dag]\hat{\rho},
\end{aligned}
\end{equation}
where we defined $\mathcal{D}[\hat{A}]\hat{\rho}=\hat{A}\hat{\rho}\hat{A}^\dag-\{\hat{A}^\dag\hat{A},\hat{\rho}\}/2$, $\kappa_\alpha$ denotes the energy damping rate associated with the bath $\alpha$, and $\hat{\rho}$ is the density matrix of the system. Here we neglect voltage fluctuations that stem from the low frequency environment\cite{gramich:2013} and we assume both cavities to be overcoupled (i.e. we neglect internal loss channels).

\begin{figure}[t!]
\centering
\includegraphics[width=.75\columnwidth]{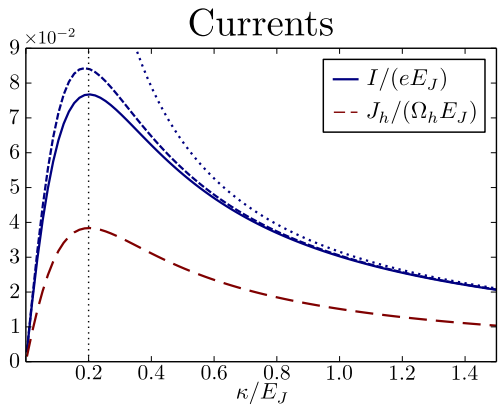}
\caption{Charge and heat currents for different ratios of $\kappa/E_J$ ($\kappa=\kappa_h=\kappa_c$). The charge current (solid, blue) is proportional to the heat current (dashed, red). For fixed $E_J$, the current shows a non-monotonic behavior as a function of $\kappa$. The toy model (dashed, blue) captures the qualitative features of the current well. Here $P(E)$ theory (dotted, blue) was employed to fit the toy model at large $\kappa/E_J$. The resulting analytic expression is in good quantitative agreement with the numerical result. The dotted line at $E_J=5\kappa$ corresponds to the performance analysis in the main text. Parameters: $n_B^h=1$, $n_B^c=0.1$, $\lambda_h=\lambda_c=0.36$, $2eV=\Omega_h-\Omega_c$, $E_J'=0.187E_J$}.
  \label{fig:current}
\end{figure}

We next focus on a voltage that satisfies the resonance condition of Eq.~\eqref{eq:voltage}, and make a RWA on our Hamiltonian, retaining only the resonant tunneling term (involving exchange of $k$ hot photons and $l$ cold photons)\cite{supplementarx}
\begin{equation}
\label{eq:hamrwa}
\hat{H}_{\rm RWA}=-\frac{E_J}{2}\bigg(i^{(l+k)}(\hat{a}_c^\dag)^l\hat{A}_c(l)\hat{A}_h(k)\hat{a}_h^k+h.c.\bigg).
\end{equation}
Here we introduced the Hermitian operators
\begin{equation}
\label{eq:aops}
\hat{A}_\alpha(k)=(2\lambda_\alpha)^ke^{-2\lambda_\alpha^2}\sum\limits_{n_\alpha=0}^{\infty}\frac{n_\alpha!}{(n_\alpha+k)!}L_{n_\alpha}^{(k)}(4\lambda_\alpha^2)|n_\alpha\rangle\langle n_\alpha |,
\end{equation}
with the generalized Laguerre polynomials $L_n^{(k)}(x)$. The validity of the RWA is discussed in the supplemental material.\cite{supplementarx}
The dc current is then computed using the RWA-form of the current operator
\begin{equation}
\label{eq:currdc}
\hat{I}_{\rm dc}=-\frac{I_c}{2i}\bigg(i^{(l+k)}(\hat{a}_c^\dag)^l\hat{A}_c(l)\hat{A}_h(k)\hat{a}_h^k-h.c.\bigg).
\end{equation}

We are interested in the power generated by the heat engine and its efficiency.
The power reads $P=IV$,
where $I=\langle \hat{I}_{\rm dc}\rangle$ and the average is taken with respect to the steady-state solution. In order to characterize the efficiency of the heat engine, we need to calculate the heat that is injected by the hot bath. To this end, we consider the time evolution of the photon number in the hot cavity\cite{walls:book}
\begin{equation}
\label{eq:dertnh}
\partial_t\langle \hat{n}_h\rangle=-i\langle[\hat{n}_h,\hat{H}_{\rm RWA}]\rangle+\kappa_h\left(n_B^h-\langle \hat{n}_h\rangle\right).
\end{equation}
The first term corresponds to the coupling to the Josephson junction while the second term corresponds to the photon flux from the heat bath. Since the photons in the hot cavity have a mean energy of $\Omega_h$, we identify the mean heat current provided by the hot bath with
\begin{equation}
\label{eq:heatcurr}
J_h=\Omega_h\kappa_h\left(n_B^h-\langle \hat{n}_h\rangle\right).
\end{equation}
In the steady state, the photon number remains constant and Eq.~\eqref{eq:dertnh} allows us to express the mean heat current in terms of the mean charge current
\begin{equation}
\label{eq:heatflux2}
\frac{J_h}{k\Omega_h}=\frac{i}{k}\langle[\hat{n}_h,\hat{H}_{\rm RWA}]\rangle=\frac{I}{2e}.
\end{equation}

Eq.~(\ref{eq:heatflux2}) implies that the heat current completely determines the charge current and vice versa. This reflects the fact that
there is no uncertainty in the energy absorbed by a tunneling Cooper pair:  every tunneling Cooper pair absorbs {\it exactly} $k$ photons from the hot bath.  This is in marked contrast
to tunneling in normal junctions. Such a proportionality between heat and charge current, also known as the strong coupling regime, has previously been discussed as a requirement to achieve high efficiencies.\cite{kedem:1965,ruokola:2012,broeck:2005,esposito:2009} We note that at resonance [i.e. for a voltage of the form given in Eq.~\eqref{eq:voltage}] the strong coupling regime is always guaranteed in our system.
Using $\eta=P/J_h$, we immediately recover Eq.~\eqref{eq:eta}. Note that we only made the RWA and assumed a steady state to obtain the universal efficiency.

\textit{Heat engine performance.} We focus on the case $k=l=1$ where we expect the highest power for experimentally achievable values of $\lambda_\alpha$. In this case, every tunneling Cooper pair is accompanied by the exchange of one photon from one cavity to the other.
For simplicity, we focus on a symmetric system with $\kappa=\kappa_h=\kappa_c$ and $\lambda=\lambda_h=\lambda_c$.

The charge and heat currents for this configuration are calculated using QuTiP\cite{johansson2013qutip} and plotted in Fig.~\ref{fig:current} as a function of $\kappa/E_J$. For small $\kappa/E_J$, the main effect of the Josephson junction is to
equilibrate the two cavities. This implies
\begin{equation}
\label{eq:occsk}
\langle\hat{n}_h\rangle=\langle\hat{n}_c\rangle=\frac{n_B^h+n_B^c}{2}+\mathcal{O}(\kappa/E_J),
\end{equation}
which is reproduced numerically (not shown).
Together with Eqs.~\eqref{eq:heatcurr} and \eqref{eq:heatflux2} this reproduces the linear behaviour of the currents for small $\kappa/E_J$. In addition to this induced equilibration, the Hamiltonian in Eq.~\eqref{eq:hamrwa} also creates coherences between the two cavities, which are an interesting subject of study in their own right.\cite{trif:2015,armour2015josephson} After reaching a maximum, the currents decrease as $1/\kappa$ at large $\kappa/E_J$.
This can be anticipated from perturbation theory (see below) and is a consequence of the energy broadened density of states in the cavities.

Figure \ref{fig:power} illustrates the power and efficiency of the system as a function of $\Omega_h/\Omega_c$ [with $V$ always tuned to the resonant value in Eq.~\eqref{eq:voltage}]. In order to act as a heat engine, the current needs to flow against the voltage bias. This is the case for frequencies in the interval $\Omega_h=[\Omega_c,T_h\Omega_c/T_c]$. The power vanishes at both endpoints of this interval. At $\Omega_h=\Omega_c$, the voltage vanishes. Since the heat can still flow from the hot to the cold bath, the efficiency also vanishes.
At $\Omega_h/T_h=\Omega_c/T_c$, the occupation numbers characterizing the baths become equal $n_B^h=n_B^c$. In this case, tunneling with the bias becomes equally probable to tunneling against the bias and both the charge as well as the heat current vanish. In this limit, the heat engine is operated reversibly and Carnot efficiency is reached. We note that this voltage is equivalent to the stopping voltage in other thermoelectric devices.

While the power thus shows a nonmonotonic behaviour as a function of $\Omega_h/\Omega_c$, the efficiency increases monotonically.
The dotted vertical lines in Figs.~\ref{fig:current} and \ref{fig:power} correspond to the realistic parameters given in table \ref{tab:params}. For these values, the heat engine produces a current of $~23\,$pA and a power of $~0.5\,$fW. This power is a factor of five higher than the power predicted in previous proposals using hybrid microwave-cavities.\cite{bergenfeldt:2014} The conversion of heat into work for these parameters happens extremely efficiently with $\eta=77.8\,\%$, which is above the Curzon-Ahlborn bound.  Both the power and the efficiency could be further increased by increasing $\Omega_h$.

The parameters in Table \ref{tab:params} were chosen as follows.  First, experimental realities constrain the maximum possible cavity frequency; here, we take $\Omega_h=2\pi\cdot13.5\,$GHz. The validity of the RWA requires $E_J \ll \Omega_c$ in order to be able to neglect non-resonant processes; we thus take $\Omega_c=10E_J$. Finally the ratio $\Omega_h/\Omega_c$ has to be chosen such that resonant processes which involve a higher number of photons can be neglected due to the smallness of the zero-point fluctuations $\lambda$. Both resonant as well as non-resonant terms that are neglected in the RWA are discussed in the supplemental material.\cite{supplementarx}

We note in passing that our setup can also be operated as a refrigerator. To this end, one has to chose the cavity frequencies such that $\langle\hat{n}_c\rangle>\langle\hat{n}_h\rangle$. A heat flux from the cold to the hot reservoir is then induced by Cooper pairs tunneling with the bias.

\textit{Toy model.} To obtain a better analytic understanding, we consider the limit of $\lambda_\alpha\rightarrow0$. Surprisingly, this limit captures the qualitative features of the heat engine even when higher order terms quantitatively modify the results. Expanding the Hamiltonian to lowest order in $\lambda_\alpha$ and focusing on the case of single photon exchange ($k=l=1$), we find
\begin{equation}
\label{eq:hamtoy}
\hat{H}_T=\frac{E'_J}{2}\left(\hat{a}_c^\dag \hat{a}_h+\hat{a}_h^\dagger\hat{a}_c\right),
\end{equation}
\begin{equation}
\label{eq:currtoy}
\hat{I}_T=-ieE_J'\left(\hat{a}_c^\dag \hat{a}_h-\hat{a}_h^\dagger\hat{a}_c\right),
\end{equation}
where $E'_J=4\lambda_h\lambda_cE_J$. This model effectively describes a particle on a tilted potential where tunneling is accompanied by the exchange of a photon [cf.~Fig.~\ref{fig:schematics}\,(b)]. In the following, we treat $E'_J$ as a fitting parameter and refer to the last equations as the toy model.

From the master equation in Eq.~\eqref{eq:master} we find\cite{walls:book}
\begin{equation}
\partial_tI_T=eE'_J\left(\langle \hat{n}_h\rangle-\langle\hat{n}_c\rangle\right)-\frac{\kappa_h+\kappa_c}{2}I_T,
\end{equation}
where $I_T=\langle \hat{I}_T\rangle$.
Using Eq.~\eqref{eq:dertnh} and an analogous equation for the cold cavity occupation number we find in the steady state (for $\kappa=\kappa_h=\kappa_c$)
\begin{equation}
\label{eq:currtavgtoykeq}
I_T=e\frac{\kappa (E'_J)^2\left(n_B^h-n_B^c\right)}{\left[\kappa^2+(E'_J)^2\right]}.
\end{equation}
The last expression, together with the fact that the heat current and the power are proportional to the charge current, provide analytical solutions for our heat engine in the case of small zero-point fluctuations $\lambda_\alpha$. Even for larger values of $\lambda_\alpha$, one obtains a good fit for the $\kappa$-dependence of quantities by treating $E'_J$ as a fitting parameter, see Fig.~\ref{fig:current}.

In the limit of $\kappa/E'_J\rightarrow 0$, $E_J'$ drops out of the last equation and it agrees with the current calculated from Eqs.~\eqref{eq:heatcurr} and \eqref{eq:heatflux2} using Eq.~\eqref{eq:occsk}. In this limit, the only effect of the Josephson junction is the complete equilibration of the two cavities, independent of other details. In the opposite limit of $\kappa/E'_J\rightarrow\infty$, the toy model reproduces the $1/\kappa$ behaviour qualitatively. To obtain the fitting parameter $E'_J$ in this limit, we employ $P(E)$ theory as described below.

\textit{P(E) Theory.} The disagreement between the toy model and numerics stems from neglected processes that arise at higher order in $\lambda_\alpha$. For large $\kappa/E_J$, these processes can be accounted for by $P(E)$-theory.\cite{NazarovPE} It expresses the current through a coherent conductor as the convolution between the tunneling rate through the conductor with $P(E)$, the probability for the environment (the two cavities here) to exchange an energy $E$ with charge carriers. Here, dissipation due to thermal baths can be accounted for through a standard input-output theory as described in Ref.~\onlinecite{NCPAT}. Note that $P(E)$-theory is a perturbative treatment in $E_{J}$ and neglects effects that appear at higher order in $E_J/\kappa$ such as equilibration of the cavities.
In the limit $\kappa/E_{J}\gg1$ (but still $\kappa\ll\Omega_h,\Omega_c$), the $P(E)$ expression for the current agrees with Eq.~\eqref{eq:currtavgtoykeq} with (here $\lambda=\lambda_h=\lambda_c$)

\begin{multline}\label{eq:fit}
	(E'_{ J})^2= E_J^2 16 \lambda^4 e^{-8 \lambda^2(1+n_{B}^{h} +n_{B}^{c} )}\\
	\sum_{m ,n=0}^{\infty}
	\tfrac{  (4\lambda^2n_{B}^{c})^m ( 4\lambda^2(1+ n_{B}^{c}))^{m} (4\lambda^2 n_{ B}^{h})^n (4\lambda^2(1+ n_{B}^{h}) )^{n}
	}{m!(m+1)! n!(n+1)!(m+n+1)}.
\end{multline}
The last expression retains the full $\lambda$-dependence of the one-photon processes relevant to resonant Cooper-pair tunneling; the complex summation reflects the interplay between zero-point fluctuations and thermal noise in the cavities.\cite{NCPAT}
Using Eq.~\eqref{eq:currtavgtoykeq} with $E_J'$ from Eq.~\eqref{eq:fit} yields a fully analytic approximation for the current which is in good quantitative agreement with the numerical calculations (cf.~Fig.~\ref{fig:current}).


\textit{Conclusions.} We have proposed and analyzed a simple yet high-power and high-efficiency mesoscopic quantum heat engine. To the best of our knowledge, our proposal constitutes the first thermoelectric heat engine where the heat current is completely separated from the electronic degrees of freedom. This is made possible by the use of a Josephson junction. The sharp energy selectivity of Cooper-pair tunneling along with the peaked spectral density of the cavities enables the conversion from heat into work with efficiencies above $75$\,\% for realistic system parameters. Our proposal is within reach of current experimental capabilities, and will lead to a better understanding of energy conversion in hybrid mesoscopic structures.

\textit{Acknowledgements.} This work was supported by NSERC.  P.P.H. acknowledges funding from the Swiss NSF.

\bibliography{biblio}

\clearpage
\widetext
\begin{center}
\textbf{\large Supplement: Quantum heat engine based on photon-assisted Cooper pair tunneling}
\end{center}

In this supplement, we derive the Hamiltonian in the rotating wave approximation (RWA) and we discuss the justification of this approximation. Throughout the supplement, equations, figures, and citations without the prefix '$S$' refer to the main text.
\setcounter{equation}{0}
\setcounter{figure}{0}
\setcounter{table}{0}
\setcounter{page}{1}
\makeatletter
\renewcommand{\theequation}{S\arabic{equation}}
\renewcommand{\thefigure}{S\arabic{figure}}

\section{\large Deriving the Hamiltonian in the RWA}

We start from the Hamiltonian given in the main text [cf.~Eq.~(3)] and transform it into a rotating frame using the unitary transformation $\hat{U}=\exp\left[{i\hat{a}_h^\dagger\hat{a}_h\Omega_ht+i\hat{a}_c^\dagger\hat{a}_c\Omega_ct}\right]$
\begin{equation}
\label{eq:rotu}
\hat{H}_{\rm R}=\hat{U}\hat{H}\hat{U}^\dagger-i\hat{U}\partial_t\hat{U}^\dagger=-\frac{E_J}{2}e^{i2eVt}\exp\left[2i\lambda_h\left(\hat{a}_h^\dag e^{i\Omega_h t}+\hat{a}_h e^{-i\Omega_h t}\right)\right]\exp\left[2i\lambda_c\left(\hat{a}_c^\dag e^{i\Omega_c t}+\hat{a}_c e^{-i\Omega_c t}\right)\right]+h.c.
\end{equation}
We then expand the Hamiltonian using the expansion of the displacement operator
\begin{equation}
\label{eq:expdispl}
\exp\left[2i\lambda_\alpha\left(\hat{a}_\alpha^\dag e^{i\Omega_\alpha t}+\hat{a}_\alpha e^{-i\Omega_\alpha t}\right)\right]=\sum\limits_{k=0}^{\infty}i^k(\hat{a}_\alpha^\dag)^k\hat{A}_\alpha(k)e^{ik\Omega_\alpha t}+\sum\limits_{k=1}^{\infty}i^k\hat{A}_\alpha(k)\hat{a}_\alpha^ke^{-ik\Omega_\alpha t},
\end{equation}
where the operators $\hat{A}(k)$ are given in Eq.~(7).
Using the above equations, setting  $2eV=k\Omega_h-l\Omega_c$, and keeping only the (time-independent) terms where $k$ hot photons are exchanged with $l$ cold photons, we recover Eq.~(6) in the main text. Note that this Hamiltonian is independent of the cavity frequencies.

\section{\large Justifying the RWA}

The RWA employed in the last section only keeps the resonant term which involves the lowest number of photons. Non-resonant as well as resonant terms which involve a higher number of photons are neglected.

All terms in the Hamiltonian which change the photon number in the hot cavity by $k$ and in the cold cavity by $-l$ fulfilling the condition $2eV=\pm(k\Omega_h-l\Omega_c)$ are resonant processes. As in the main text, we focus on a voltage tuned to $2eV=\Omega_h-\Omega_c$. The only term which is kept in the main text corresponds to the exchange of a single photon, i.e. $k=\pm1$ and $l=\pm1$. For $\Omega_h=4.5\Omega_c$, the next resonant term in terms of the number of photons involved corresponds to $k=\pm1$ and $l\pm8$, i.e. one hot cavity photon is exchanged with $8$ cold cavity photons. The corresponding term in the Hamiltonian reads
\begin{equation}
\hat{H}'=-\frac{E_J}{2}\left(i\hat{a}_h^\dagger\hat{A}_h(1)\hat{A}_c(8)\hat{a}_c^8+h.c.\right).
\end{equation}
The current operator then obtains the additional term
\begin{equation}
\hat{I}'=-\frac{I_c}{2i}\left(i\hat{a}_h^\dagger\hat{A}_h(1)\hat{A}_c(8)\hat{a}_c^8-h.c.\right).
\end{equation}
Including these terms in the numerical calculation leads to changes in the current of the order of $10^{-5}$\,pA. The above terms can thus safely be neglected. This is due to the proportionality $\hat{A}_\alpha(k)\propto\lambda_\alpha^k$ which surpresses a process that involves nine photons by a factor of $\lambda^9\approx10^{-4}$.
All other neglected resonant processes involve an even higher number of photons and are thus suppressed even more.

For our choice of frequencies, the smallest detuning for any non-resonant term is $\Omega_c/2$. The term in the Hamiltonian involving the smallest number of photons with this detuning corresponds to $k=0$ and $l=\pm3$. This process corresponds to a tunneling Cooper pair which absorbs/emits three cold cavity photons. The detuning of this term is given by the mismatch in energy $2eV-3\Omega_c=\Omega_c/2$. The corresponding term in the Hamiltonian reads
\begin{equation}
\hat{H}_\Omega(t)=\frac{E_J}{2}\left(i\hat{A}_h(0)\hat{A}_c(3)\hat{a}_c^3e^{i\frac{\Omega_c}{2}t}+h.c.\right)=\hat{F}e^{i\frac{\Omega_c}{2}t}+h.c.
\end{equation}
An estimate of the importance of this term can be given by comparing the relevant matrix elements of $\hat{F}$ to the detuning $\Omega_c/2$. We find
\begin{equation}
\frac{\sqrt{\langle \hat{F}\hat{F}^\dagger+\hat{F}^\dagger\hat{F}\rangle}}{\Omega_c/2}\approx 0.0092,
\end{equation}
where the expectation value is taken with respect to the RWA solution.

We conclude that only keeping the resonant term which involves the smallest number of photons is justified due to the large detuning and/or the smallness of the matrix elements of the neglected terms.

\end{document}